\documentclass[a4paper]{article}

\usepackage{INTERSPEECH2022}
\usepackage{booktabs}
\usepackage{multirow}
\usepackage{amssymb} 
\usepackage{pifont} 
\newcommand{\cmark}{\ding{51}} 
\newcommand{\xmark}{\ding{55}} 
\usepackage{cite}
\usepackage{color}

\title{Coarse-to-Fine Recursive Speech Separation for Unknown Number of Speakers}
\name{Zhenhao Jin$^1$, Xiang Hao$^2$, Xiangdong Su$^{1*}$}
\address{
  $^1$ College of Computer Science, Inner Mongolia Univeristy, Hohhot, China\\
  $^2$ The Chinese University of Hong Kong}
\email{cssxd@imu.edu.cn}

\begin{document}

\maketitle
\begin{abstract}
The vast majority of speech separation methods assume that the number of speakers is known in advance, hence they are specific to the number of speakers. By contrast, a more realistic and challenging task is to separate a mixture in which the number of speakers is unknown. This paper formulates the speech separation with the unknown number of speakers as a multi-pass source extraction problem and proposes a coarse-to-fine recursive speech separation method. This method comprises two stages, namely, recursive cue extraction and target speaker extraction. The recursive cue extraction stage determines how many computational iterations need to be performed and outputs a coarse cue speech by monitoring statistics in the mixture. As the number of recursive iterations increases, the accumulation of distortion eventually comes into the extracted speech and reminder. Therefore, in the second stage, we use a target speaker extraction network to extract a fine speech based on the coarse target cue and the original distortionless mixture. Experiments show that the proposed method archived state-of-the-art performance on the WSJ0 dataset with a different number of speakers. Furthermore, it generalizes well to an unseen large number of speakers.
\end{abstract}


\section{Introduction}

Spontaneous conversations often contain overlapping speech with multiple active speakers. This phenomenon, called cocktail party~\cite{haykin2005cocktail}, has been studied for several decades~\cite{wang2018supervised}. Despite the recent advances in deep learning technologies for various speech signal processing tasks, e.g., speech recognition~\cite{qian2018single,seki2018purely,von2020multi} and speaker identification~\cite{abdullah2021sok,li2020bridging,horiguchi2020end}, speech separation remains unresolved. Speech separation still represents a major source of error within the whole speech signal processing system. 

Most speech separation methods can only deal with mixtures with a fixed number of speakers and the speaker number must be estimated in advance. For instance, only a prior speaker number is estimated, methods based on permutation invariant training (PIT)~\cite{yu2017permutation,kolbaek2017multitalker,yousefi2019probabilistic} can pool over all possible permutations for all sources and use the permutation with the lowest error to update the network. For the methods based on Deep clustering (DC)~\cite{hershey2016deep,isik2016single,wang2018alternative,luo2017deep}, a pretrained neural network is firstly used to generate discriminative embedding for each time-frequency (T-F) bin. Then by giving a correct number of clusters equal to the true number of speakers, these embedding vectors are clustered by some algorithms~\cite{makino2007blind} to obtain desired sources. In summary, most previous works assume that the number of speakers in the mixture is prior. However, in reality, the number of speakers is unknown in advance and the number of speakers can vary from two to more than five~\cite{von2019all,kinoshita2020tackling}.

To separate mixture with variable number speakers, some separation methods integrating explicit or implicit speaker counting are proposed~\cite{wang2021count, kanda2020joint, nachmani2020voice, chazan2021single}. ~\cite{wang2021count, kanda2020joint} use an explicit speaker counting module to count the number of speakers on the frame level or utterance level. Then a speaker separation model based on PIT is utilized to separate multiple speakers. ~\cite{nachmani2020voice,chazan2021single} use a voice activity detector~\cite{yang2010comparative} and a different model for every number of speakers. ~\cite{zhu2021multi} trains a single model with a count-head to infer the number of speakers and multiple decoder heads to separate the mixture.
Besides, the methods based on a recursive manner also can separate mixtures with an unknown number of speakers~\cite{takahashi2019recursive, von2020multi}. Instead of separating all speakers in a mixture at once, these methods separate only one speaker from a mixture at a time and the residual speech is fed back to the separation model for the recursion to separate the next speaker. However, the main drawback of these methods is that the distortion in the extracted speech and residual speech is getting starker and starker as the iterations increase. In other words, the methods based on a recursive manner are not suitable to deal with a mixture with a large number of speakers.

\begin{figure*}[th]
  \centering
  \includegraphics[width=\linewidth]{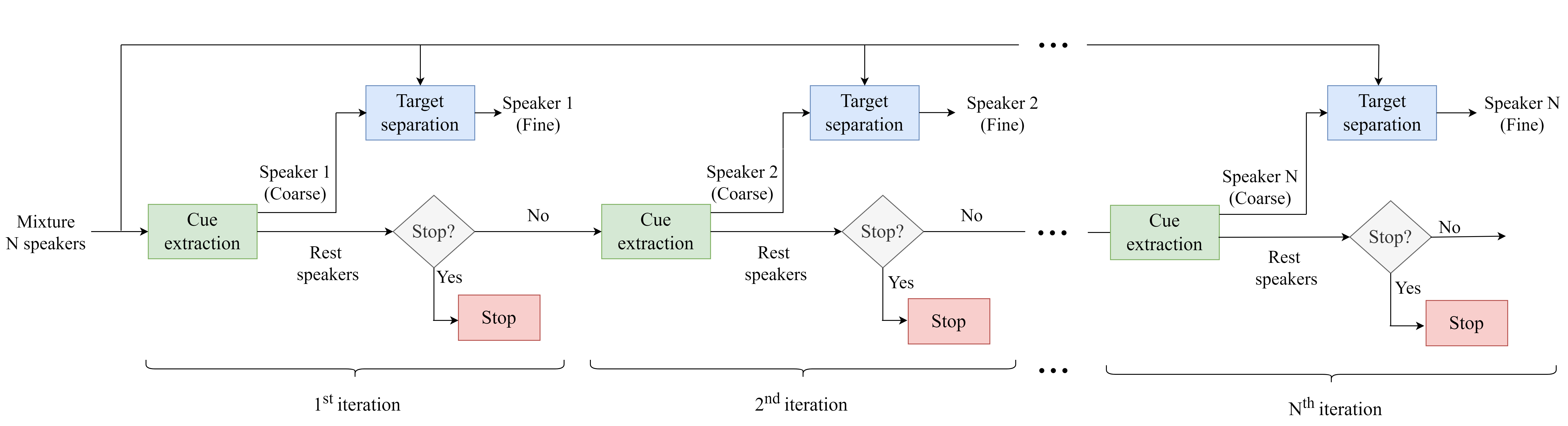}
  \caption{Workflow of the coarse-to-fine recursive speech separation method.}
  \label{image1}
  \vspace{-.4cm}
\end{figure*}

Considering the merits and limitations of the above methods, we cast the problem as a recursive multi-pass source extraction problem and propose a coarse-to-fine recursive speech separation method. Specifically, the proposed method contains two stages from coarse to fine, namely, recursive cue extraction and target speaker extraction. In the recursive cue extraction stage, a Dual-Path RNN (DPRNN)~\cite{luo2020dual} is used recursively to learn and determine how many iterations have to be performed. It outputs a cue speech source and the residual at a time from the mixture. After that, it will repeat this process until all speech sources are extracted.
To make the separation model not sensitive to the distortion of the recursive manner, we further propose a second stage, target speaker extraction. To be specific, the extracted speech from the first stage is treated as a coarse-grained target speech. Then, we use it as the cue of the target speech to extract fine-grained target speech from the original mixture. In this way, we can avoid the separation degradation caused by multiple recursive separations as much as possible. It is worth noting that the manner in the second stage is very similar to that in the conventional target speaker extraction tasks, whose effectiveness has been proven extensively~\cite{delcroix2019compact,he2020speakerfilter,xu2020spex}. Compared to the conventional target speech extraction tasks, we only use the target speech extracted from the first stage instead of an offline registered speech. In other words, the extracted cue speech is in the same acoustic environment as that of the mixture, hence the needed length of cue speech is far shorter. Upon combining the above two stages, the proposed method can determine how many computational iterations have to be performed and avoid the accumulation of distortion that comes into the extracted speech in each iteration.


\section{Method}

There are two stages in the proposed method. As shown in Figure~\ref{image1}, the first stage uses a recursive cue extraction model to extract one target speech from a mixture as a cue at a time. Moreover, the model will automatically determine whether the recursive separation needs to repeat or stop. The second stage takes the mixture and the cue speech as the input of the target speech separation model to generate the speaker’s pure speech.

\subsection{Recursive cue extraction}
 The conventional recursive speech separation models separate each speaker's speech from the mixture iteratively and regard the separated speech as the final result. In these models, the remaining mixture from the current iteration is used as input for the next iteration. Since input distortion accumulates with each iteration, the quality of the sequentially separated speeches will decrease with the increase of iterations. This paper proposes a coarse-to-fine two-stage separating method to deal with the problem of distortion accumulation. Different from the recursive speech separation model, our method first employs a recursive cue extraction model to extract the coarse target cue and then separates each target speech from the original mixture with the help of the corresponding coarse cue. We use the DPRNN~\cite{luo2020dual} as our recursive cue extraction model because of its strong capabilities in possessing time-series data, and train it as~\cite{takahashi2019recursive}. There are two output channels of the recursive cue extraction model. The first output channel is the coarse target cue of one speaker, and the second channel is the rest mixture. Except for the first iteration in which the input of the recursive cue extraction model is the original mixture, the input of the other iterations comes from the second output channel of the last recursive separation.We define the unknown number of speakers as $N$ and the cue speech of $N$ speakers as $c_{1}(t), c_{2}(t), \dots, c_{N}(t)$ from a mixture signal $x(t)$, where $x(t)=\sum_{i=1}^{N}c_{i}(t)$. For the $j$th recursion step, the recursive cue extraction model separates one speaker cue speech $\hat{c}^{j}(t)$ and the mixture of residual speaker speech $\hat{r}^{j}(t)$ from $\hat{r}^{j-1}(t)$ as
\begin{equation}
\label{function1}
\hat{c}^{j}(t), \hat{r}^{j}(t) = F(\hat{r}^{j-1}(t))
\end{equation}
where $F(\cdot)$ represents the recursive cue extraction model. To avoid exceptions, we define $\hat{r}^{0}(t)=x(t)$. To extract the $N$ speakers, the recursive cue extraction model will run $N$ times. We use the stop condition in~\cite{takahashi2019recursive}, which has 95.7\% accuracy in repeat-and-stop binary classifier.

\subsection{Target speech separation}

\begin{figure}[t]
  \centering
  \includegraphics[width=.85\linewidth]{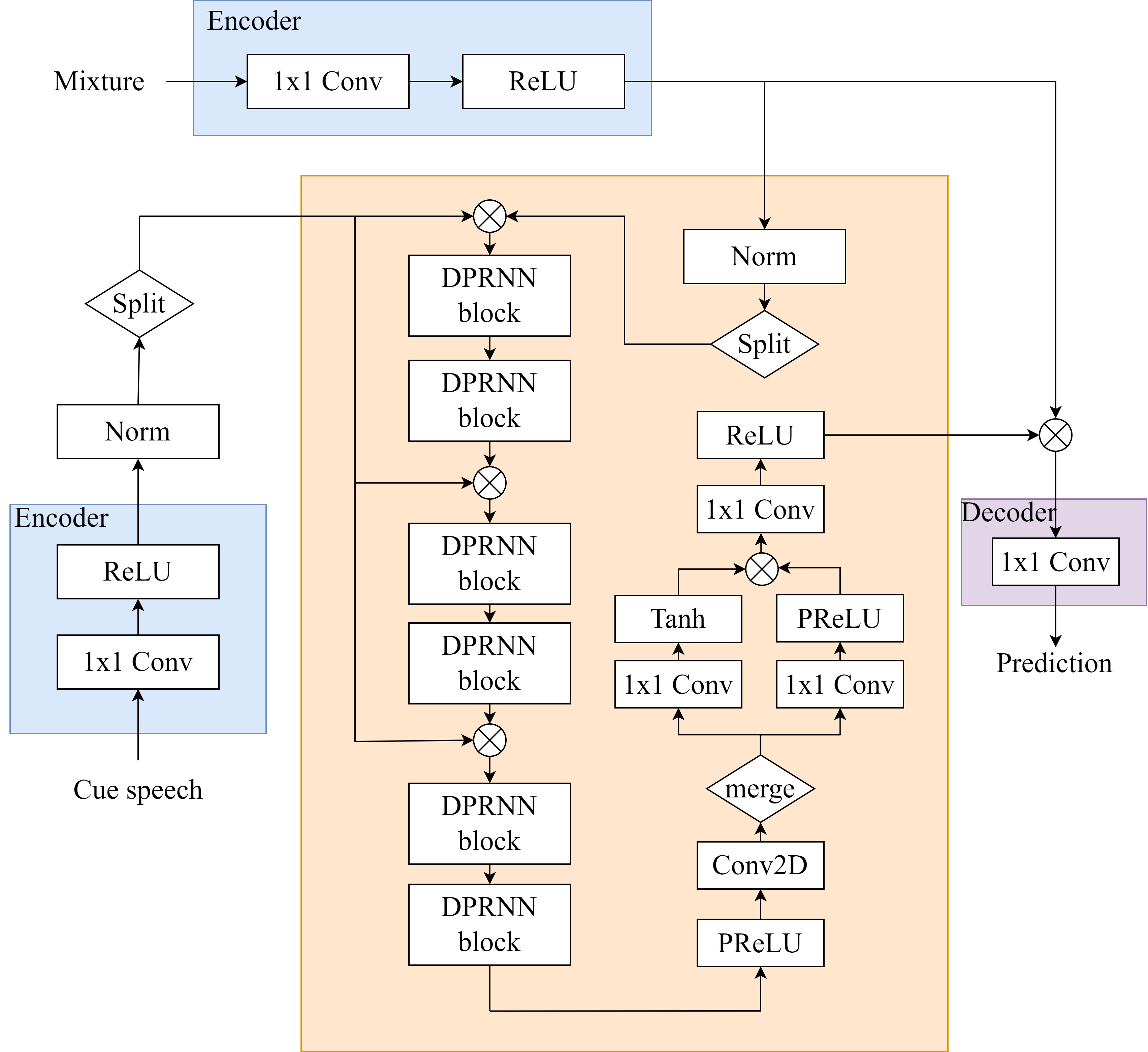}
  \caption{Workflow of the target speech separation based on the DPRNN blocks.}
  \label{image2}
  \vspace{-.6cm}
\end{figure}

After the previous processing stage, we get the cue speech of each target speaker. In the target speech separation stage, we use the coarse target cues from the first stage to facilitate the fine target speech extraction. Specifically, we extract the fine target speech from the mixture instead of extracting it from the output of the previous recursive separation. It can avoid the input distortion caused by recursive separation as much as possible. In the first stage, we use the encoder-decoder speech separation structure whose backbone is also the DPRNN in the second stage. It excludes interference factors of difference in model performance. As shown in Figure~\ref{image2}, the inputs of the target speech separation model include a mixture and a coarse target cue. The single channel output is the fine target speech corresponding to the coarse target cues. Similar to the DPRNN separation module, we use the encoder which consists of pointwise convolution (1×1-conv) and rectified linear unit (ReLU) to encode the mixture and the coarse target cue. To make the target separation model convergence during training, we use two different encoders for coarse target cue and mixture, respectively. The encoder of coarse target cue comes from the first-stage model's encoder, especially. But the encoder of the mixture is a trainable new one. After encoding the mixture and coarse target cues, we normalize them and split them as operating in~\cite{luo2020dual}. We use the dot product to guide the target speech separation, and it has happened when the split feature needs to as input at the odd number DPRNN block, where the main module of the target speech separation has six DPRNN blocks. We replace the softmax non-linearity, which dot products with another gate to generate the masks, with PReLU non-linearity similarity as~\cite{takahashi2019recursive}. The mask multiplies mixed speech features to obtain the features of the fine target speech. Finally, we use the decoder to generate time-domain fine target speech. Corresponding to the first stage mentioned the $j$th recursion step, the target separation can be defined as:
\begin{equation}
\label{function2}
\hat{o}^{j}(t) = G(x(t), \hat{c}^{j}(t))
\end{equation}
where $G(\cdot)$ is target separation model. The $\hat{o}^{j}(t)$ is the prediction of target speech corresponding to $\hat{c}^{j}(t)$. Because each separation in the second stage has input from the first stage. So it also separates N times same as the first stage. But it does not separate 2N times in all of the proposed method because the proposed method is a complete recursive speech separation process.

\subsection{Loss function}
As described in Section 2.2, we use the One-and-Rest PIT (OR-PIT)~\cite{takahashi2019recursive} as the loss function of the first stage. The OR-PIT needs to calculate the loss between the first channel and the $N$ possible speakers and the loss between the second channel and the remaining channels. The OR-PIT is defined as:
\begin{equation}
\label{function3}
L = \mathop{\min}\limits_{i}l(\hat{c}(t),c_{i}(t))+\frac{1}{N-1}l(\hat{r}(t),\sum_{n \neq i}c_{n}(t)) .
\end{equation}
where $l$ is a scale-invariant source-to-noise ratio (SI-SNR) loss function~\cite{luo2020dual,luo2019conv}, which is define as 
\begin{equation}\label{pb1}
\left \{
\begin{array}{l}
    s_{\text{target}}:=\frac{\left \langle \hat{s},s \right \rangle s}{ \left \| s \right \|^2} \\
    e_{\text{noise}}:= \hat{s} - s_{\text{target}} \\
    l_{\text{SI-SNR}}(\hat s,s):= 10 {\log}_{10}\frac{ \left \| s_{\text{target}} \right \|^2}{ \left \| e_{\text{noise}} \right\|^2} \\
\end{array}
\right.
\end{equation}
where $\hat{s}$ and $s$ are the mean normalized estimates and targets, respectively. The mean normalization of the sources ensures the scale invariance property of the loss function. In the second stage, we directly use the SI-SNR as the loss function since there is no permutation problem in this stage.

\section {Experiments}

\subsection{Datasets and metrics}
To evaluate the proposed method, we build the datasets based on the Wall Street Journal dataset (WSJ0) corpus \cite{garofolo1993csr}. The open-source MatLab code for simulating mixture with overlapped speakers can be found in GitHub\footnote{https://github.com/JunzheJosephZhu/MultiDecoder-DPRNN/tree/master/tools/2345mix}, which includes WSJ0-2mix (two speakers), WSJ0-3mix (three speakers)~\cite{hershey2016deep}, WSJ0-4mix (four speakers), WSJ0-5mix (five speakers)~\cite{nachmani2020voice}. Each of these four datasets contains three parts, namely, the training set, validation set, and test set. The training set and the validation set are created by using the utterances in the ``si\_tr\_s" dataset. The WSJ0-mix were randomly chosen and combined with a random SNR value between 0 to 5dB. The test datasets are created using the si\_et\_s and si\_dt\_s with 16 speakers, who differ from those speakers in the training set.
We use SI-SNR to measure the quality of the separated speech, which has been extensively used in speech separation~\cite{luo2020dual,luo2019conv}.

\subsection{Training details}
\label{sec:training_details}

The sampling rate of all speech data is fixed to 8,000 Hz. We segment the mixture and the target speech into four-second lengths with a two-second overlap (50\%). Then, we pad the two-to-four-second length speech to a four-second length speech with zero. For the speech with a length that is less than two seconds, they will be discarded. 

We randomly combined the mixture in WSJ0-2mix and WSJ0-3mix datasets to train our first-stage model. After the first-stage training, we fixed the cue-extraction model and began to train the second-stage separation model. We just use the WSJ0-3mix to train our second-stage separation model, which is trained three times for iterative separation. Both two models used the same settings. The initial learning rate is $5 \times 10^{-4}$. When the SI-SNR does not decline five epochs on the validation set, the learning rate is reduced by half. The gradient clipping with a maximum L2-norm of 5 is applied during training. We use the Adam~\cite{kingma2015adam} optimizer with momentum parameters (0.9, 0.999) to control the optimization during the training. We used the best model configuration of the DPRNN described in~\cite{luo2020dual}. 
We also fine-tune the first-stage model using the manner described in~\cite{takahashi2019recursive}. 
In particular, to fine-tune the first stage, we replace WSJ0-2mix with a two-speaker mixture obtained from the separation of the first iteration on WSJ0-3mix.


\section{Results and discussions}

\begin{table}[t]
\setlength\tabcolsep{5pt}
\caption{Performance on the different number of speakers after adding the cue-based target speaker extraction. The SI-SNR (in dB) on WSJ0-2mix, WSJ0-3mix, WSJ0-4mix and WSJ0-5mix are compared.}
\label{tab:usage_stage_1}
\centering
\begin{tabular}{@{}cccccccc@{}}
\toprule
\multirow{2}{*}{ID} & \multirow{2}{*}{\#spk} & \multicolumn{2}{c}{System} & \multicolumn{4}{c}{Number of speakers}                \\ \cmidrule(l){3-8} 
            &           & Stage 1      & Stage 2     & 2            & 3            & 4          & 5          \\ \midrule
1 & \multirow{2}{*}{2}     & \cmark            & \xmark           & 18.4            &-2.1     &-3.3    &-6.6            \\
2 &                       & \cmark            & \cmark           & 20.1         &-1.7      & -3.8    &-7.1            \\ \midrule
3 & \multirow{2}{*}{2 \& 3}   & \cmark            & \xmark           & 17.9 & 12.9 & 8.3 & 3.9 \\
4 &                       & \cmark            & \cmark           & 18.6        & 13.9        & 9.8       & 5.6       \\ 
                       \bottomrule       
\end{tabular}
\vspace{-.4cm}
\end{table}

\subsection{Effectiveness of the target speech extraction stage}
Table~\ref{tab:usage_stage_1} shows an ablation experiment for evaluating the effectiveness of the second stage, i.e., target speaker extraction guided by the cue extracted by the first stage. Firstly, we trained the proposed model with and without the target speaker extraction stage on WSJ0-2mix (two speakers). The experimental results correspond to ID 1 and ID 2, respectively. Compared to models with and without the second stage (ID 1 vs. ID 2), the model's performance improves. This result shows that the second stage is helpful. One thing that should be noticed is that the SI-SNR on datasets with three, four, and five speakers are negative. That's because the model trained only on a two-speaker dataset does not generalize well to an unseen number of speakers.
Furthermore, we mixed WSJ0-2mix and WSJ0-3mix datasets as a new training dataset. The corresponding results are shown in ID 3 and ID 4. Still, after adding the second stage, the model's performance is better than the single first stage. It even improves 1.7 dB on the WSJ0-5mix test dataset. In addition, we can find that using a training dataset with more speakers will enhance the generalization of a large number of unseen speakers. In summary, the second stage model can further use the coarse target cue to extract fine target speech with better quality.

\begin{table}[t]
\setlength\tabcolsep{5pt}
\caption{Performance on the different number of speakers after adding the cue-based target speaker extraction based on the fine-tuned first stage (``w/o FT": First stage without fine-tuning. ``w/ FT": First Stage with fine-tuning).}
\label{tab:usage_stage_2}
\centering
\begin{tabular}{@{}cccccccc@{}}
\toprule
\multirow{2}{*}{ID} & \multirow{2}{*}{\#spk} & \multicolumn{2}{c}{System} & \multicolumn{4}{c}{Number of speakers}                \\ \cmidrule(l){3-8} 
& & Stage 1      & Stage 2     & 2            & 3            & 4          & 5          \\ \midrule
1 & 2 \& 3 & \cmark            & \xmark           &17.9             &12.9         &8.3     & 3.9           \\
2& (w/o FT) & \cmark            & \cmark           & 18.6            &13.9        &9.8    &  5.6          \\ \midrule
3 & 2 \& 3   & \cmark            & \xmark           & 17.8 & 13.7 & 9.9 & 6.2 \\
4 & (w/ FT)                   & \cmark            & \cmark           & 19.8        & 15.4        & 11.4       & 7.3 \\  \bottomrule  
\end{tabular}
\vspace{-.2cm}
\end{table}

\subsection{Effectiveness of fine-tuning speaker cue extraction}
As described in Section~\ref{sec:training_details}, we fine-tuned the first-stage model on WSJ0-2mix and WSJ0-3mix training datasets. Table~\ref{tab:usage_stage_2} shows the performance of the first-stage model and the two-stage model after fine-tuning. ``w/ FT" and ``w/o FT" represent the models with and without fine-tuning, respectively. 
The results show that fine-tuning can improve the performance of the single first-stage model on almost all test datasets (ID 1 vs. ID 3). In addition, the performance of the two-stage model also shows a significant improvement based on the fine-tuned first-stage model (ID 2 vs. ID 4). This phenomenon indicates that the first-stage model extracts more fine-grained cue speech after fine-tuning. Moreover, these fine-grained cue speech can further improve the performance of the second stage.

\begin{table}[t]
\setlength\tabcolsep{4.5pt}
\caption{Performance on each number of iterative separations }
\label{tab:usage_stage_3}
\centering
\begin{tabular}{@{}ccccccccc@{}}
\toprule
\multirow{2}{*}{ID} & \#spk & \multicolumn{2}{c}{System} & \multicolumn{5}{c}{Iteration times of separation} \\ \cmidrule(l){3-9} 
 & (Test) & Stage 1      & Stage 2     & 1            & 2            & 3          & 4       & 5   \\ \midrule
 1 & \multirow{2}{*}{2}     & \cmark            & \xmark           &18.9              &16.7              &-    &-   & -        \\
 2 &                     & \cmark            & \cmark           &19.9              &19.7              & -      & -    &  -     \\ \midrule
3 &\multirow{2}{*}{3}   & \cmark            & \xmark           &14.7  &14.5  &12.0  & -  &-\\
4 &                       & \cmark            & \cmark           &15.6         &15.4         &15.2        &  -    & - \\ \midrule
5 & \multirow{2}{*}{4} & \cmark            & \xmark           &11.5    &10.3   &10.1       &7.7     &-  \\
6 &                       & \cmark            & \cmark           & 12.4     &12.0   &11.7   &9.5            & - \\ 
                     \midrule
7 & \multirow{2}{*}{5} & \cmark            & \xmark           &8.8     &6.6      &6.2   &5.9      &3.4  \\
8 &   & \cmark       & \cmark      &9.0      &7.8      &7.0     &6.7   &5.7  \\   \bottomrule       
\end{tabular}
\vspace{-.5cm}
\end{table}

\subsection{Generalization to a large number of unseen speakers}
Table~\ref{tab:usage_stage_3} further shows the performance of the proposed method for each iteration on all test datasets. The single first-stage models (ID 1,3,5,7) and the two-stage models (ID 2,4,6,8) correspond to the models of ID 3 and ID 4 in Table~\ref{tab:usage_stage_2}, respectively. In Table~\ref{tab:usage_stage_3}, ID 1 represents the single first-stage model (ID 3 in Table~\ref{tab:usage_stage_2}) evaluated on WSJ0-2mix (two speakers) test datasets. Therefore, to separate all speakers, this model have to perform two iterations. ID 2 is similar to ID 1, but it uses the two-stage model (ID 4 in Table~\ref{tab:usage_stage_2}). ID 3,4,5,6,7,8 are similar to ID 1,2, but they corresponds to test datasets with more speakers. 
From Table~\ref{tab:usage_stage_3}, we can find that adding the second stage will improve the performance of the model regardless of the number of speakers and iterations. In addition, as the number of iterations increases, the quality of the speech separated from either the one-stage model or the two-stage model decreases.
However, it is obvious that the quality decay of the two-stage model is smaller. That's because the two-stage model treats the separated speech from the first stage as a cue, which is then used to extract the fine-grained speech from the original distortionless mixture.


\begin{table}[t]
  \caption{Comparison with baselines on the WSJ0-2mix, WSJ0-3mix, WSJ0-4mix, and WSJ0-5mix datasets. (``-": Results are unavailable. ``$1$": Implementation of Conv-TasNet-gLN in~\cite{takahashi2019recursive}. ``$2$": Implementation of Conv-TasNet-gLN in~\cite{nachmani2020voice})}
  \label{tab:baseline}
  \centering
  \begin{tabular}{ccccc}
    \toprule
    \multirow{2}{*}{System} & \multicolumn{4}{c}{Number of speakers}
    \\ \cmidrule(l){2-5}
      &\textbf{2}  & \textbf{3}    & \textbf{4}    & \textbf{5}   
    \\
    \midrule
    Ground-truth                 & 0.0      & -3.1      & -5.0       & -6.2  \\
    Conv-TasNet-gLN$^{1}$~\cite{takahashi2019recursive}           & 14.6      & 11.6      & -       & -  \\
    Conv-TasNet-gLN$^{2}$~\cite{nachmani2020voice}           & 15.3      & 12.7      & 8.5       & 6.8   \\
    DPRNN~\cite{nachmani2020voice}                     & 18.8      & 14.7      & 10.4     & \textbf{8.7}  \\
    Voicerint (Conv-TasNet)~\cite{hao2020unified}   & 16.0      & -      & -       & -   \\
    Voiceprint (DPRNN)~\cite{hao2020unified}         & 17.4      & -      & -       & -  \\
    Conv-TasNet + OR-PIT~\cite{takahashi2019recursive}       & 14.8      & 12.6      & 10.2       & -   \\
    DPRNN + OR-PIT               & 17.8      & 13.7      & 9.9       & 6.2  \\
    \textbf{Ours}                & \textbf{19.8}      & \textbf{15.4}     & \textbf{11.4}      & 7.3   \\
    \bottomrule
  \end{tabular}
  \vspace{-.6cm}
\end{table}

\subsection{Comparison with baselines}
As shown in Table~\ref{tab:baseline}, we compare the following six baseline methods: Conv-TasNet~\cite{luo2019conv} is a time-domain model with a convolution structure for speech separation. DPRNN~\cite{luo2020dual} is a time-domain model with LSTM structure for speech separation. Voiceprint with Conv-TasNet and Voiceprint with DPRNN~\cite{hao2020unified}, are target speaker extraction models based on Conv-TasNet and DPRNN models. Conv-TasNet with OR-PIT~\cite{takahashi2019recursive}, which is a recursive speech separation model trained with Conv-TasNet. Similarly to Conv-TasNet with OR-PIT, we implement the DPRNN with OR-PIT. It should notice that the Conv-TasNet-gLN $^{1}$ is the implementation of~\cite{takahashi2019recursive}. The Conv-TasNet-gLN $^{2}$ is the implementation of~\cite{nachmani2020voice}. The result is shown in Table~\ref{tab:baseline}. Each column depicts a different dataset, where the number of speakers $N$ in the mixed-signal $x(t)$ is different. As can be seen, the proposed method is superior to previous methods by a sizable margin, in three datasets except for five speakers dataset. Compared with time-domain separation models, our approach can deal with the variable unknown number of speakers' speech separation. The proposed method uses shorter reference but gets better performance than the voiceprint-based method. Compared with the methods based on OR-PIT, our approach can avoid the input distortion accumulation to use the coarse target cue separating fine target speech. It is worth noting that the performance of our method in five-speaker is not the best. The performance of DPRNN with OR-PIT in five-speaker is also poor. It proves the quality of coarse target cues can influence the second stage model to extract fine target speech.

\section{Conclusion}
In this paper, we proposed a coarse-to-fine recursive speech separation method for the unknown number of speakers. Our method includes two stages. The first stage can extract the coarse target cue recursively from the mixture. The second stage inputs the coarse target cue and mixture to extract the fine target speech. We compared our approach with recursive speech separation models on the WSJ0-mix test dataset. The results showed that our approach achieves state-of-the-art performance. We also found that the proposed two-stage model decreases the distortion accumulation of input and has a better generalization to an unseen large number of speakers.

\tiny
\bibliographystyle{IEEEtran}
\bibliography{newbib}
\end{document}